\documentclass[pra,twocolumn,showpacs,aps,floatfix,amsmath,amssymb,amsfonts,superscriptaddress]{revtex4-1}
\usepackage{graphicx}
\usepackage{amsmath,amssymb,revsymb}
\usepackage{color}
\usepackage[colorlinks=true,citecolor=blue]{hyperref}
\usepackage{braket}
\begin{document}

\title{Effective-range approximations for resonant scattering of cold atoms}

\author{Caroline L. Blackley}
\affiliation{Joint Quantum Centre (JQC) Durham/Newcastle, Department of
Chemistry, Durham University, South Road, Durham, DH1~3LE, United Kingdom}
\author{Paul S. Julienne}
\affiliation{Joint Quantum Institute, University of Maryland and NIST, College
Park, Maryland, 20742, USA}

\author{Jeremy M. Hutson}
\affiliation{Joint Quantum Centre (JQC) Durham/Newcastle, Department of
Chemistry, Durham University, South Road, Durham, DH1~3LE, United Kingdom}
\begin{abstract}
Studies of cold atom collisions and few-body interactions often require the
energy dependence of the scattering phase shift, which is usually expressed in
terms of an effective-range expansion. We use accurate coupled-channel
calculations on $^{6}$Li, $^{39}$K and $^{133}$Cs to explore the behavior of
the effective range in the vicinity of both broad and narrow Feshbach
resonances. We show that commonly used expressions for the effective range
break down dramatically for narrow resonances and near the zero-crossings of
broad resonances. We present an alternative parametrization of the effective
range that is accurate through both the pole and the zero-crossing for both
broad and narrow resonances. However, the effective range expansion can still
fail at quite low collision energies, particularly around narrow resonances. We
demonstrate that an analytical form of an energy and magnetic field-dependent
phase shift, based on multichannel quantum defect theory, gives accurate
results for the energy-dependent scattering length.
\end{abstract}
\date{\today}
\maketitle

\section{Introduction} \label{sec:introduction}

The study of trapped samples of ultracold atomic gases is an extremely fruitful
area of experimental and theoretical research.  It includes studies of
Bose-Einstein condensation (BEC) of bosonic
species~\cite{Inouye1998,Cornish2000,Cornell2002,Ketterle2002}, the crossover
between the BEC and Bardeen-Cooper-Schrieffer regimes of fermionic
species~\cite{Bourdel2004,Zwierlein2004,Varenna2006}, the production of
ultracold polar molecules~\cite{Jochim2003,Ni:KRb:2008,Lang:ground:2008}, the
manipulation of atoms in optical lattices~\cite{Greiner2008,Bloch2005}, and the
study of Efimov physics in few-body
systems~\cite{Wang:2011:NJP,Roy2013,Ferlaino2011,Berninger2011}.  The theory of
such phenomena has been greatly simplified by the ability to characterize the
zero-energy interaction of two atoms in terms of the $s$-wave scattering length
$a$.  For many species, nearly any desirable value $a(B)$ can be obtained by
tuning a magnetic field $B$ near the pole position $B_0$ of a threshold
scattering resonance known as a Feshbach resonance. The scattering length is
approximately related to the magnetic field by the formula~\cite{Moerdijk:1995}
\begin{equation}
    a(B) = a_{\rm bg} \left ( 1 - \frac{\Delta}{B-B_0} \right ) , \label{aB}
\end{equation}
where $\Delta$ is the width of the resonance and  $a_{\rm bg}$ is the
background scattering length far from resonance.

The parametrization of low-energy interactions in terms of $a(B)$ allows the
detailed chemical interaction between two ultracold atoms in the limit of zero
collision kinetic energy $E \to 0$ to be replaced by a zero-range Fermi
pseudo-potential whose strength is proportional to $a(B)$.    However, as
experimental probes of ultracold systems become more powerful and
sophisticated, the variation of atomic interactions as a function of energy
away from exactly $E=0$ must be considered and understood. The usual way to
describe the variation with energy of the near-threshold $s$-wave scattering
phase shift $\eta(E)$ is to use an effective-range expansion at small collision
momentum $\hbar k$, where $E=\hbar^2 k^2/(2\mu)$ and $\mu$ is the reduced mass
of the two atoms~\cite{Bethe:1949,Hinckelmann:1971},
\begin{equation}
 k \cot{\eta(E)} = -\frac{1}{a_0} + \frac12 r_{\rm eff} k^2 + \ldots\, , \label{r0}
\end{equation}
where the parameter $r_{\rm eff}$ is called the effective range and $a_0$ is
the zero-energy scattering length.  We prefer a modified way of writing this
expression and define the energy-dependent scattering length $a(E)$ by
\cite{Hutson:res:2007},
\begin{equation}
 a(E) = -\frac{\tan{\eta(E)}}{k} = \frac{1}{ik} \frac{1-S(E)}{1+S(E)} , \label{ak}
\end{equation}
where $S=e^{2i\eta}$ is the diagonal element of the unitary S-matrix for the
threshold channel in question.  With this formulation, both $\eta(E)$ and
$a(E)$ are real when only elastic scattering is possible but become complex in
the presence of inelasticity. Eq.~\eqref{r0} becomes
\begin{subequations}
\begin{align}
  a(E)^{-1} = a_0^{-1} - \frac12 r_{\rm eff} k^2 + \ldots\, , \label{r0alt} \\
\intertext{ or}
  a(E) = a_0 + \frac12 r_{\rm eff} a_0^2 k^2 + \ldots\, . \label{r0alt2}
\end{align}
\end{subequations}
Far from a pole or a zero-crossing in $a_0(B)$, finite-difference equations
based on either of these relationships may be used to evaluate $r_{\rm eff}$.
However, those based on \eqref{r0alt} are numerically unstable near a
zero-crossing and those based on \eqref{r0alt2} are numerically unstable near a
pole.

Effective-range expansions have been invoked to include the role of collisions
at finite energy in few-body phenomena~\cite{Braaten2006, Blume:rpp:2012,
WangRev2013, Wang2011d, Dyke2013} and to correct for the zero-point energy in
optical lattice physics~\cite{Naidon2007}.  The energy-variation of the phase
shift is needed to obtain the contribution of two-body collisions to low-energy
partition functions and thermodynamic properties of cold
gases~\cite{Mies:1982}.  The effective range is known to vary around Feshbach
resonances \cite{Zinner:2009, Gao:QDT:1998, Flambaum1999, Gao2011a, Dyke2013,
Roy2013}, but there has been no in-depth numerical study of the behavior of
$\eta(E)$, $a(E)$ and $r_{\rm eff}$ as $B$ is tuned across Feshbach resonances
of different types. In the present work we use accurate coupled-channels
calculations to explore this numerically for both broad and narrow Feshbach
resonances. Our calculations demonstrate that the effective-range expansion can
fail in some circumstances for low-energy atomic collisions and also elucidate
the range of applicability of simple approximations that have been developed to
relate the effective range to the scattering length, given the form of the
long-range potential \cite{Julienne:2006}. We also present an approach based on
multichannel quantum defect theory (MQDT) \cite{Julienne:2006}, which gives an
analytic form for the energy-dependence of the phase shift that applies even
when the effective-range expansion breaks down. We will demonstrate that this
analytic representation gives excellent agreement with coupled-channels
calculations for both broad and narrow resonances.

We choose to study resonances in $^{6}$Li, $^{133}$Cs and $^{39}$K, in their
lowest possible $s$-wave collision channels, all of which are important in
studies of Efimov physics \cite{Naidon:2011:6Li, Braaten:2010:6Li,
Ottenstein:2008:6Li, Huckans:2009:6Li, Zaccanti:2009:39K, Roy2013,
Ferlaino2011, Kraemer:2006, Berninger2011}. The interaction potentials used in
the coupled-channels calculations are those of Z\"urn {\em et al.}
\cite{Zurn:Li2-binding:2013} for $^6$Li, Berninger {\em et al.}\
\cite{Berninger2011, Berninger:Cs2:2013} for $^{133}$Cs, and Falke {\em et
al.}\ \cite{Falke:2008} for $^{39}$K. The atomic hyperfine/Zeeman states are
labelled using Roman letters a,b,c, etc., in increasing order of energy.

The structure of the paper is as follows: Section II describes effective-range
theory, including the magnetic field dependence of the effective range; Section
III discusses the limits of the effective-range expansion; Section IV describes
the MQDT approach to the energy-dependent scattering length; Section V details
the effectiveness of this new approach and Section VI concludes with a summary
of the applicability and accuracy of both effective-range theory and the
MQDT-based approach.

\begin{figure*}[!htbp]
\includegraphics[width=\textwidth]{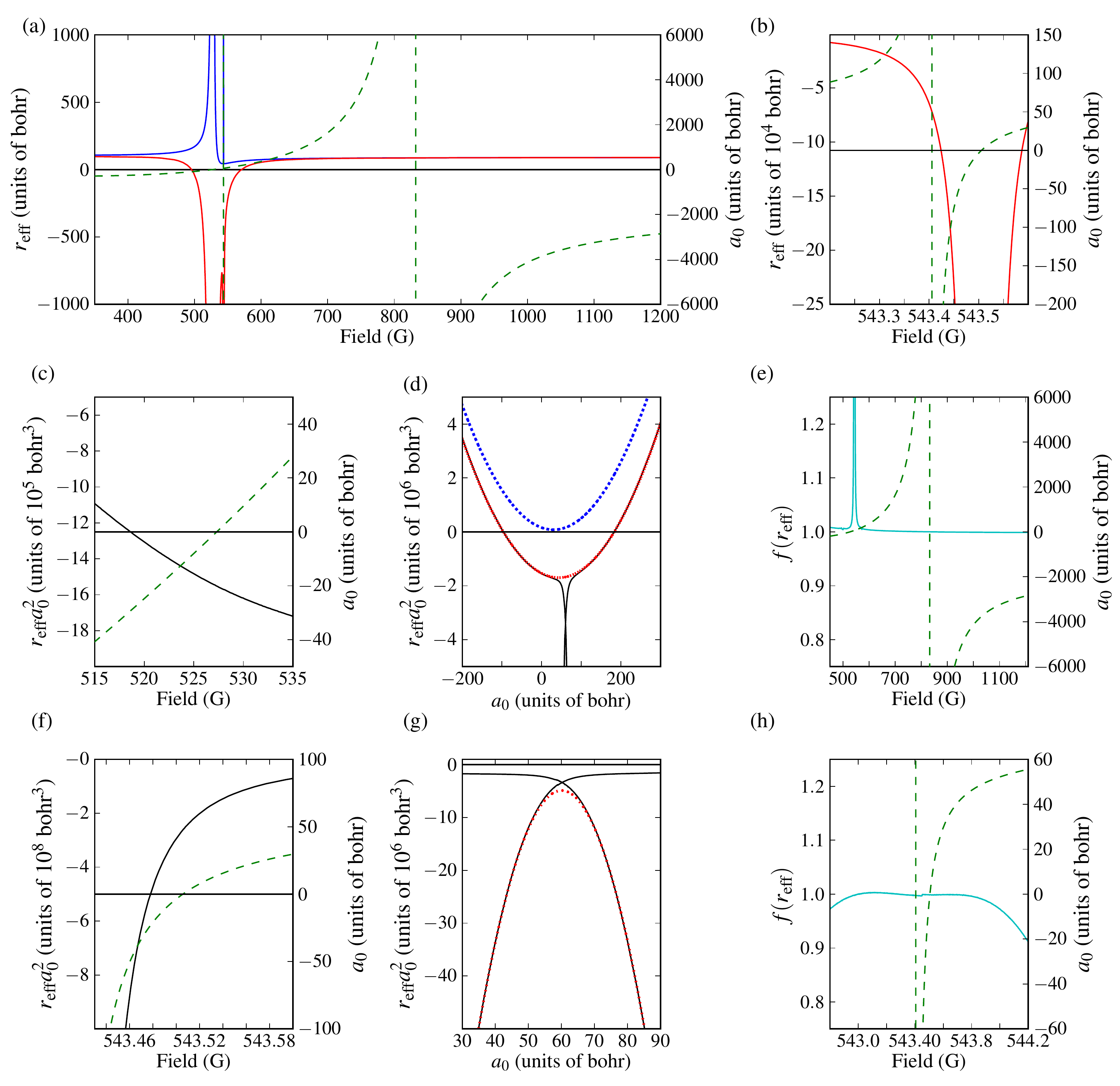} \caption{(a) The
field-dependent effective range for the ab channel of $^{6}$Li from
coupled-channel calculations (red solid or light gray) and as calculated from
Eq.\ \eqref{r0gao} (blue solid or dark gray). The zero-energy s-wave scattering
length is also shown (green dashed). (b) An expanded view of (a) showing the
narrow resonance at $543.40$~G. (c) and (f) The quantity $r_{\rm eff}a_0^2$
(black solid), which is a smoothly varying function of magnetic field through
the zero-crossing of $a_0$ (green dashed) for both the wide resonance (c) and
the narrow resonance (f). (d) The quantity $r_{\rm eff}a_0^2$ (black), which is
parabolic as a function of $a_0$ across the width of the wide resonance, except
around the narrow resonance. The red dotted line shows the parabola
$v+r_0(a_0-a_{\rm ext})^2$ fitted to the coupled-channel results, while the
blue dotted (upper) line shows the corresponding parabola from Eq.\
\eqref{r0gao2}. (e) The function $f(r_{\rm eff})$ of Eq.\ \eqref{freff} (cyan),
with parameters appropriate for the wide resonance, which is constant across
the width of the wide resonance in $a_0$ (green dashed) except around the
narrow resonance. (g) the quantity $r_{\rm eff}a_0^2$ (black solid), which is
parabolic as a function of $a_0$ across the width of the narrow resonance; the
red dotted line shows the parabola $v+r_0(a_0-a_{\rm ext})^2$ fitted to the
coupled-channel results. (h) The function $f(r_{\rm eff})$ of Eq.\
\eqref{freff} (cyan), with parameters appropriate for the narrow resonance,
which is constant across the width of the narrow resonance in $a_0$ (green
dashed).} \label{fig:r0-func}
\end{figure*}

\section{Behavior of the effective range near a Feshbach resonance}
\label{sec:eff_range_intro}

In this section we analyse the behavior of the effective range in the vicinity
of Feshbach resonances of different types. A magnetically tunable resonance can
be classified as broad or narrow, based on the parameter $s_{\rm res}$
\cite{Chin:RMP:2010},
\begin{equation}
s_{\rm res} =
\frac{a_{\rm bg}}{\bar{a}}\frac{\delta\mu \Delta}{\bar{E}} , \label{sres}
\end{equation}
where $a_{\rm bg}$ is the ``local" background scattering length, $\delta\mu$ is
the difference between the magnetic moment of the bare bound state and the
magnetic moment of the separated atoms, and $\Delta$ is the width of the
resonance. The length and energy scales $\bar{a}$ and $\bar{E}$ are as defined
by Gribakin and Flambaum \cite{Gribakin:1993},
\begin{eqnarray}
\bar{a}&=&\frac{2\pi}{\Gamma(1/4)^2}
\left( \frac{2\mu C_6}{\hbar^2}\right)^{1/4} \label{abar} \\
\bar{E}&=&\frac{\hbar^2}{2\mu \bar{a}^2}. \label{ebar}
\end{eqnarray}
Using these scalings allow us to define dimensionless length and energy
parameters, ${a}/\bar{a}$ and ${E}/\bar{E}$, respectively. Resonances with
$s_{\rm res} > 1 $ are referred to as broad resonances, whilst those with
$s_{\rm res} < 1 $ are narrow resonances.

The effective-range expansion is the leading term in a Taylor series and breaks
down at `high' energies. However, in the present work it is always valid up to
at least $E/k_{\rm B}=50$~nK. We therefore obtain $r_{\rm eff}$ at each
magnetic field by performing coupled-channels calculations at 1~pK and 10~nK
and fitting the resulting values of $a(E)$ from Eq.\ \eqref{ak} to either Eq.\
\eqref{r0alt} or Eq.\ \eqref{r0alt2}. The coupled-channels calculations are
performed using the {\sc molscat} package \cite{molscat:v14}, adapted to handle
collisions in external fields \cite{Hutson:field:2011}. Calculations are
carried out with a fixed-step log-derivative propagator
\cite{Manolopoulos:1986} at short range and a variable-step Airy propagator
\cite{Alexander:1984} at long range. The wavefunctions are matched to their
long-range solutions, the Ricatti-Bessel functions, to find the S-matrix
elements; these are related to the energy-dependent scattering length and phase
shift by Eq.\ \eqref{ak}.

Gao~\cite{Gao:QDT:1998} and Flambaum {\it et al.}~\cite{Flambaum1999} have
developed an approximate formula relating $r_{\rm eff}$ to $a$, based on the
case of single-channel scattering with an $R^{-6}$ potential,
\begin{equation}
r_{\rm eff} \approx \left(\frac{\Gamma(1/4)^4}{6\pi^2}\right)\bar{a}
\left[1-2\left(\frac{\bar{a}}{a_0}\right)
+2\left(\frac{\bar{a}}{a_0}\right)^2\right] . \label{r0gao}
\end{equation}
We show below that this formula works well near the pole of a broad resonance,
but may break down around a zero-crossing. In particular, \eqref{r0gao}
predicts that $r_{\rm eff}$ is always positive, which is not in fact the case.
For narrow resonances, we demonstrate that the parabolic dependence on $1/a_0$
is retained, but quite different coefficients are required.

To contrast the behavior of the effective range across broad and narrow
resonances, we consider $^6$Li in its lowest (ab) $s$-wave scattering channel.
Using an $L=0$ ($s$-only) basis set, the scattering length for this channel has
only two resonances at fields below 1000~G, one broad near 832~G
($\Delta=-262$~G) and the other narrow near 543.40~G ($\Delta=0.10$~G)
\cite{Zurn:Li2-binding:2013}. The system is somewhat unusual because the narrow
resonance is close to the zero-crossing of the broad resonance. However, as the
spacing in magnetic field between the two features is several orders of
magnitude greater than the width of the narrow resonance, the overall behavior
of the two features is still distinct.

The scattering length $a(B)$ and effective range $r_{\rm eff}(B)$ for $^6$Li
are shown in Figure \ref{fig:r0-func}(a) between 200 and 1000~G. For the wide
resonance, the effective range is a smooth function of magnetic field except
near the zero-crossing in $a_0(B)$ close to 527~G, where it diverges to
negative values. This may be contrasted with the behavior of Gao's formula
\eqref{r0gao}, also shown in Fig.\ \ref{fig:r0-func}, which diverges to
positive values. The quantity $r_{\rm eff}a_0^2$, shown in Fig.\
\ref{fig:r0-func}(c) as a function of field, is continuous through the
zero-crossing, but naturally diverges at the resonance pole, where $r_{\rm
eff}$ itself does not. As shown in Fig.\ \ref{fig:r0-func}(d), it is close to
parabolic as a function of $a_0$, except close to the narrow resonance.
However, the parabola dips below zero between $a_0=183$ and $-96$ bohr,
accounting for the fact that $r_{\rm eff}$ is negative in this region. The
corresponding parabola from Gao's formula,
\begin{equation}
r_{\rm eff}a_0^2 \approx \left(\frac{\Gamma(1/4)^4}{6\pi^2}\right)
\left[\bar{a}^3+\bar{a}(a_0-\bar{a})^2\right], \label{r0gao2}
\end{equation}
is also shown in Fig.\ \ref{fig:r0-func}(d). It is similar to the true parabola
but is offset from it and is positive everywhere, with a minimum value of
$\bar{a}^3\Gamma(1/4)^4/6\pi^2=77840$ bohr$^3$ at $a_0=\bar{a}$.

\begin{table*}
\begin{tabular}{cccccccccc}
\hline
\hline
\quad System &\quad $\bar{a}$~(bohr)&\quad $B_0$~(G)&\quad $\Delta$~(G) &\quad$a_{\rm bg}$~(bohr)& \quad $s_{\rm res}$&\quad $r_0$~(bohr) &\quad $a_{\rm ext}$~(bohr) &\quad $v$ (bohr$^3$) &\quad $-2R^*$ (bohr) \\
\hline
$^{6}$Li		&$29.88$	&\multicolumn{2}{c}\hfil Eq.\ \eqref{r0gao}\hfil	&$-$			&$-$					&$87.19$		&$29.88$		&$77840$				&$-$		\\
$^{6}$Li		&$29.88$	&$832$	&$-262$		&$-1593$		&$27$				&$87$	&$43$	&$-1.7\times10^6$	&$-1.1$	\\
$^{6}$Li		&$29.88$	&$543.40$	&$0.10$	&$59.0$		&$8.1\times10^{-4}$	&$-71000$&$60$	&$-4.9\times10^6$	&$-74000$	\\
$^{39}$K		&$61.77$	&$744.93$	&$-0.005$	&$-33.4$		&$6.2\times10^{-4}$	&$-190000$&$-34$	&$2.2\times10^6$	&$-200000$\\
$^{133}$Cs	&$96.62$	&$226.73$	&$0.076$		&$2062$		&$0.19$				&$-810$	&$2800$	&$1.8\times10^9$	&$-1000$ \\
\hline
\hline
\end{tabular}
\caption{Parameters of Eq.\ \eqref{freff} that characterize $r_{\rm eff}$ in
the vicinity of resonances of different types.  \label{tab:rstar_tab}}
\end{table*}

The values of $r_{\rm eff}a_0^2$ from coupled-channel calculations may be
fitted to a parabola \begin{equation} r_{\rm eff}a_0^2=v+r_0(a_0-a_{\rm
ext})^2, \label{parab} \end{equation} with parameters given in Table
\ref{tab:rstar_tab}. By construction, $r_0$ is the value of $r_{\rm eff}$ at
the resonance pole. The quantity
\begin{equation}
f(r_{\rm eff}) = \frac{r_{\rm eff}a_0^2}{v+r_0(a_0-a_{\rm ext})^2} \label{freff}
\end{equation}
is almost constant across the whole width of the broad resonance, except close
to the narrow resonance, as shown in Fig.\ \ref{fig:r0-func}(e).

In the narrow-resonance region, the effective range varies very fast with
magnetic field even very close to the pole, as shown in Fig.\
\ref{fig:r0-func}(b). An expanded view of $r_{\rm eff}a_0^2$ in this region is
shown in Fig.\ \ref{fig:r0-func}(g). It is clear that $r_{\rm eff}a_0^2$ is
actually double-valued as a function of $a_0$: the narrow resonance contributes
a second near-parabolic feature, but it has completely different parameters
from the parabola for the broad resonance.  We have fitted a parabola of the
same form to points away from the region around $a_0=59$~bohr, where the narrow
resonance reaches its background scattering length and rejoins with the wide
resonance, and the resulting curve is shown in Fig.\ \ref{fig:r0-func}(g). The
parameters of the parabola for the narrow resonance, also given in Table
\ref{tab:rstar_tab}, bear no resemblance to those from Gao's formula
\eqref{r0gao2}.

Petrov \cite{Petrov:narrow:2004} and Bruun {\em et al.} \cite{Bruun:2005}
introduced a parameter $R^*$, defined as
\begin{equation}
 R^* = \frac{\hbar^2}{2\mu a_{\rm bg} \Delta \delta\mu}
 =\frac{\bar{a}}{s_{\rm res}}. \label{rstar}
\end{equation}
For narrow resonances, this is large and positive and is related to the
effective range at the pole by $R^*\approx -r_0/2$. The values obtained from
this expression are included in Table \ref{tab:rstar_tab}; it may be seen that
$R^*$ is within about 4\% of $-r_0/2$ for the narrow resonance in $^6$Li, but
(as expected) bears no resemblance to it for the broad resonance.

To explore further the behavior of the effective range around narrow
resonances, we have carried out additional calculations on the resonances at
744.93~G in the aa channel of $^{39}$K and at 226.73~G in the aa channel of Cs.
The $^{39}$K resonance is caused by an $L=0$ bound state, whereas the Cs
resonance is caused by an $L=2$ bound state. The quantity $r_{\rm eff}a_0^2$
was again found to be close to parabolic in each case, with parameters given in
Table \ref{tab:rstar_tab}. It may be seen that $r_0$ and $v$ may have the same
or different signs; when they are different, $r_{\rm eff}$ diverges at the
zero-crossing with the opposite sign to its value at the pole. For narrow
resonances, the position of the extremum in $r_{\rm eff}a_0^2$, $a_{\rm ext}$,
is typically close to $a_{\rm bg}$. This is consistent with the expression
given by Zinner and Thogerson \cite{Zinner:2009} for $r_{\rm eff}$ in the
vicinity of a narrow resonance, $r_{\rm eff}=r_0(1-a_{\rm bg}/a_0)^2$. However,
this expression gives $r_{\rm eff}=0$ far from resonance. If we add a
``background" effective range $r_{\rm eff,bg}$, the resulting parabola for
$r_{\rm eff}a_0^2$ is of the form of Eq.\ \eqref{parab}, with
\begin{subequations}
\begin{align}
  a_{\rm ext}&=a_{\rm bg}(1-r_{\rm eff,bg}/r_0)  \\
\intertext{ and}
  v&=a_{\rm bg} a_{\rm ext} r_{\rm eff,bg}.
\end{align} \label{aextv}%
\end{subequations}
For resonances that are not very narrow, this effect can make $a_{\rm ext}$
significantly different from $a_{\rm bg}$, as seen for the Cs resonance at
226.73~G in Table \ref{tab:rstar_tab}. For the isolated resonances in $^{39}$K
and Cs, Gao's formula \eqref{r0gao}, evaluated for $a_0=a_{\rm bg}$, gives
$r_{\rm eff,bg}$ and hence $v$ and $a_{\rm ext}$ within 10\% of the values in
Table \ref{tab:rstar_tab}. Eq.\ \eqref{parab}, together with parameters from
Eqs.\ \eqref{r0gao}, \eqref{rstar} and \eqref{aextv}, thus provides a useful
approximate expression for $r_{\rm eff}$ in the vicinity of an isolated narrow
resonance that does not require coupled-channel calculations.

\section{Limitations of the effective-range expansion} \label{sec:eff_range_analysis}

\begin{figure*}[!htbp]
\centering
\includegraphics[width=1\textwidth]{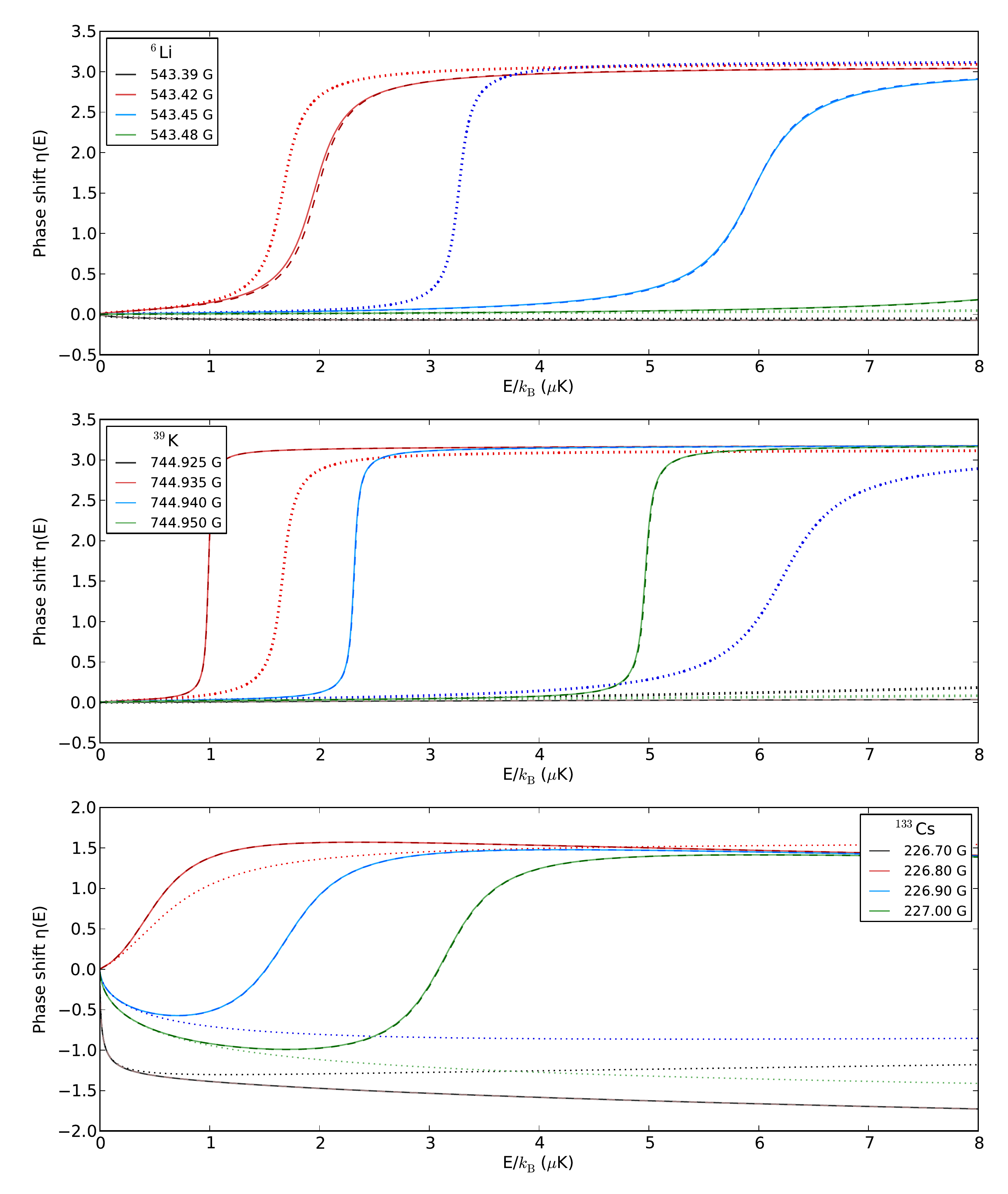}
\caption{Energy dependence of the phase shift $\eta(E)$ at magnetic fields
around narrow resonances in $^{6}$Li, $^{39}$K and $^{133}$Cs. Coupled-channels
calculations (solid lines) give the actual variation. The effective-range
expansions (dotted lines) are given by Eq.\ \eqref{r0}, except for those at
543.48 G for $^6$Li, 744.925 G for $^{39}$K, and 226.80 G for $^{133}$Cs, which
are close to zeroes in $a_0(B)$ and are therefore calculated with the
phase-shift form of Eq.\ \eqref{r0alt2}.  The effective-range expansions
deviate substantially from the coupled-channels results at collision energies
on the order of  $\mu$K. The MQDT approach of Section \ref{sec:eta_intro}
(dashed lines) gives an accurate representation of $\eta(E)$ over the full
range of collision energies. Top: The resonance at $B_0=543.40$~G in the ab
channel of $^{6}$Li; center: The resonance at $B_0=744.93$~G in the aa channel
of $^{39}$K; bottom: The resonance at $B_0=226.73$~G in the aa channel of
$^{133}$Cs.} \label{fig:phase_shift}
\end{figure*}

In this section, we assess the range of energies over which the effective-range
expansion provides an accurate representation of the energy-dependent
scattering length. We consider the same 4 resonances as in Section
\ref{sec:eff_range_intro}, at collision energies ranging from 1~nK to 1~mK. For
each resonance, we calculate the energy-dependent phase shift at multiple
magnetic fields around the zero-energy pole position. For resonances at the
lowest atomic threshold, the state responsible for the resonance is always
bound on the low-field side of the zero-energy resonance pole. We therefore
calculate $\eta(E)$ at one field just below the pole and several fields above
it.

\begin{table}[tbp]
\begin{tabular}{cccc}
\hline
\hline
&\quad $B$ (G) &\quad $a_0(B)$ (bohr)  & \quad $r_{\rm eff}$ (bohr) \\
\hline
$^6$Li&$ 543.39$	&$456.0$&$-5.4\times10^4$\\
  &$ 543.42$		&$-362.5$&$-9.7\times10^4$	\\
  &$ 543.45$		&$-75.3$&$-2.3\times10^5$\\
  &$ 543.48$		&$-20.8$&$-1.1\times10^6$	\\
\hline
$^{133}$Cs & $226.7$  &7680.6		&$-290$	\\
  &226.8& $-103.4$	&$-4.6\times10^5$ \\
  &226.9 & 	1152.2	&$ -420 $	\\
  &227.0 	& 1485.6	&$ 86 $\\
\hline
$^{39}$K	&$744.925$	& -4.6	&$-7.0\times10^7$\\
  &	 744.935&$ -82.0$&$-6.6\times10^4$	\\
  &	744.940& $-54.1$ &$-2.7\times10^4$	\\	
  &744.950	& $-43.1$	&$-8000$			\\
\hline
\hline
\end{tabular}
\caption{Parameters for effective-range calculations. At each magnetic field,
the zero-energy scattering length is given along with the effective range as
calculated using the effective-range expansion.} \label{tab:reff_fail_tab}
\end{table}

Fig.\ \ref{fig:phase_shift} compares the energy-dependent phase shift directly
from coupled-channels calculations with that from the effective-range
expansion, Eq.\ \eqref{r0}, using the accurate (field-dependent) values of
$r_{\rm eff}$ from the previous section. The values of the effective range at
the specific fields shown are given in Table \ref{tab:reff_fail_tab}.
Significant deviations can be seen for energies on the order of $1~\mu$K. For
Cs at $B=226.80$~G, for example, the effective-range expansion is inadequate at
energies above 200~nK, corresponding to $E/\bar{E}= 2 \times 10^{-3}$. On the
high-field side of the pole, there is a quasibound state at low collision
energy; as the energy passes through this, the phase shift $\eta$ increases by
$\pi$, and there is a pole in the energy-dependent phase shift $a(B)$ when
$\eta(E)=\pi/2$; the location of this feature is not well captured by the
effective-range expansion. This is particularly true for the Cs resonance at
226.73~G, where the non-resonant part of the phase shift has a general
downwards trend as a function of energy.

\begin{figure}[tbp]
\centering
\includegraphics[width=\columnwidth]{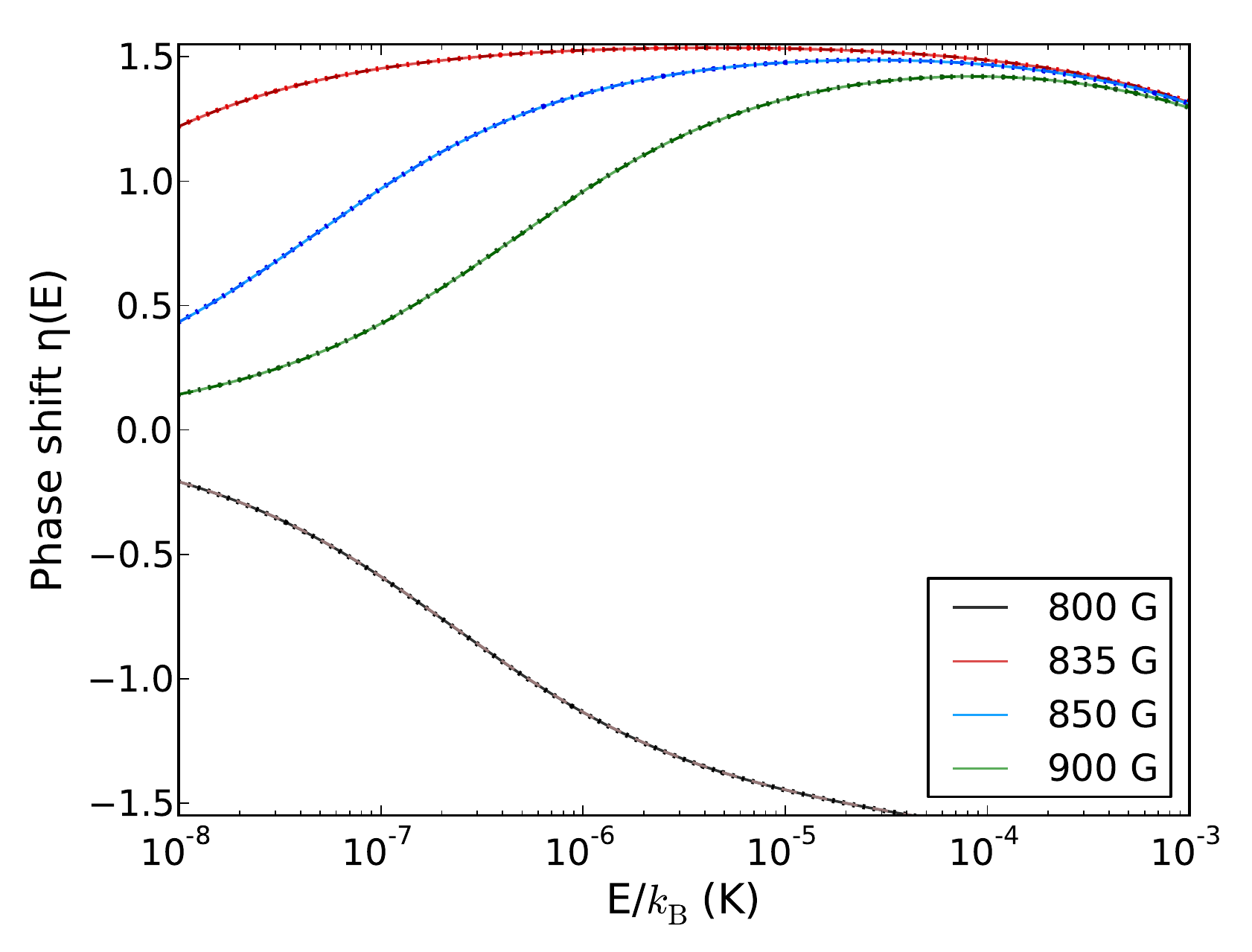}
\caption{Energy dependence of the phase shift, $\eta(E)$, at magnetic fields
around the broad resonance in $^{6}$Li. The results of coupled-channels
calculations, the effective-range expansion and the MQDT approach are
indistinguishable on this scale.} \label{fig:broad_reff}
\end{figure}

We have also analysed the broad s-state resonance in $^{6}$Li at 832~G, which
has $s_{\rm res} \gg 1$, and the results are shown in Fig.\
\ref{fig:broad_reff}. In this case the effective-range expansion is
indistinguishable from the results from coupled-channels calculations.

\section{MQDT approach to an energy-dependent phase shift} \label{sec:eta_intro}

A more complete theory of the energy dependence of the phase shift may be
formulated in the framework of Multichannel Quantum Defect Theory (MQDT).
Julienne and Gao \cite{Julienne:2006} have described a two-channel MQDT
approach to resonant scattering of ultracold atoms, combining the MQDT approach
of Julienne and Mies \cite{Mies:1984a,Mies:1984} with the analytic van der
Waals theory of Gao \cite{Gao:QDT:1998, Gao:2000, Gao:2001, Gao:2004}. A
similar theory has been described by Gao \cite{Gao:2008}, but in quite
different notation.

\begin{figure}[!tbp]
\includegraphics[width=1\columnwidth]{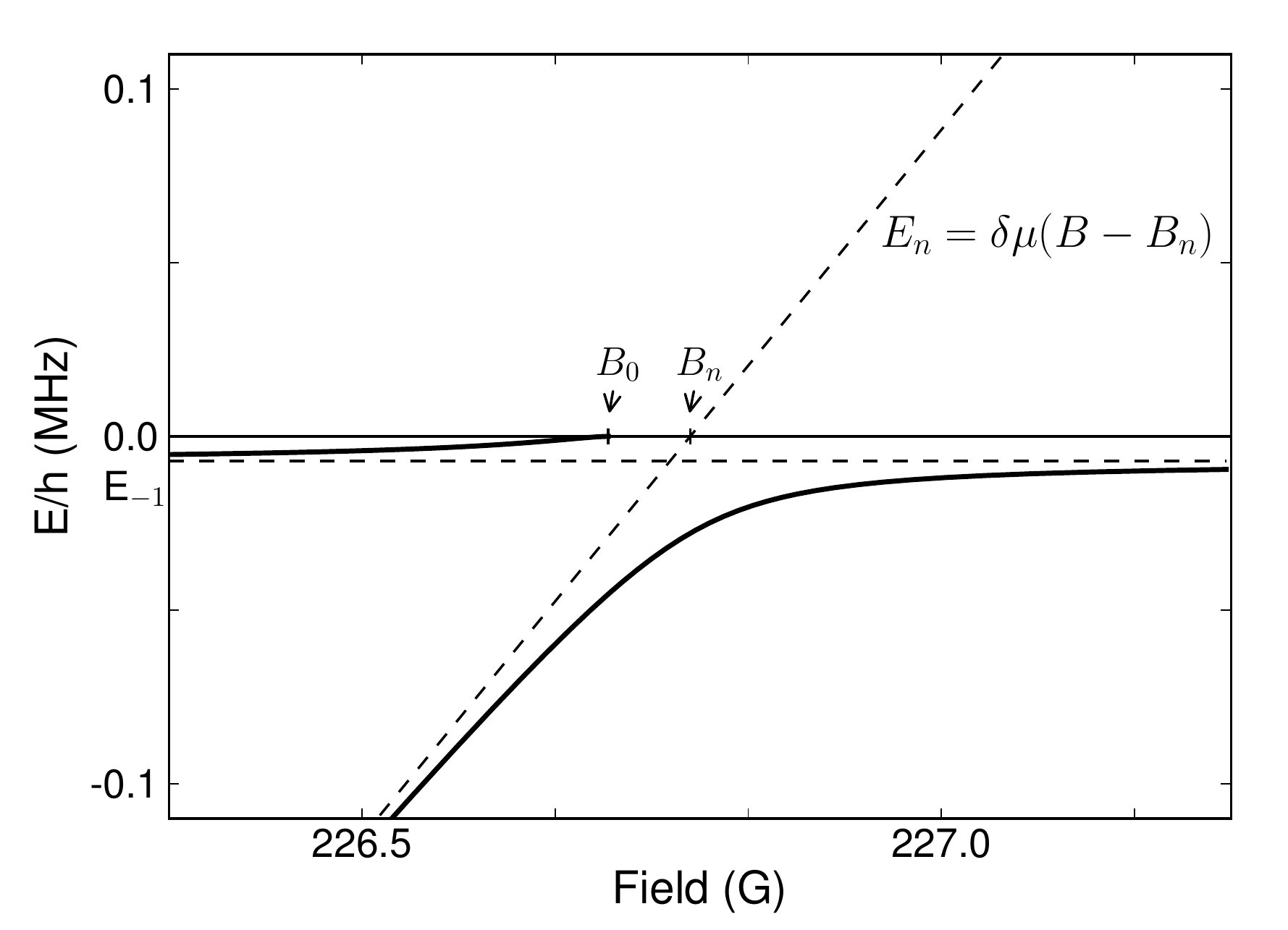}
\caption{The resonance at 226.73~G in the aa channel of Cs. The two-channel model
(dashed lines) includes a bare bound state which crosses threshold at $B_n$ and
a reference potential whose first bound level is at $E_{-1}$.  The bound states
from the coupled-channels calculations (solid lines) have have an avoided
crossing with the resonance pole at $B_0$.} \label{fig:twochan_model}
\end{figure}

For an isolated resonance, the complex set of many coupled channels can be
approximated by a two-channel model where the closed channel is represented by
a `bare' bound state with energy $E_n$ and the open channel by a `bare'
continuum state characterized by the background scattering length of the
resonance. The key quantities are illustrated in Fig.\ \ref{fig:twochan_model}
for the $^{133}$Cs resonance near 226.73~G. The energy of the bare bound state
with respect to threshold is $E_n= \delta\mu(B-B_n)$. The phase shift at fixed
magnetic field follows the Breit-Wigner form, $\eta(E) = \eta_{\rm bg}+
\eta_{\rm res}(E)$, where $\eta_{\rm bg}$ is the background component and
$\eta_{\rm res}$ is the resonant component,
\begin{equation}
\eta_{\rm res}(E)=-\tan^{-1}\left(\frac{\frac12\Gamma_n}{E-E_0}\right).
\end{equation}
Here $\Gamma_n$ is the resonance width and the resonance position $E_0$ differs
from $E_n$ by a shift $\delta E_n$, with $E_0=E_n+\delta E_n$. Near threshold,
$\eta_{\rm bg}(E)$, $\Gamma_n(E)$ and $\delta E_n(E)$ are strongly
energy-dependent and their functional forms may be obtained from MQDT.

MQDT connects the energy-insensitive short-range potential to the
energy-sensitive long-range part of the potential, using the solutions for a
reference potential that closely resembles the true potential at long range.
The solutions for the reference potential are given at short range by
WKB-normalized wavefunctions and at long range by asymptotic Bessel functions.
However, at energies near threshold the WKB description breaks down at long
range and the short-range solutions are connected to the long-range solutions
using the MQDT functions $C(E)$ and $\tan\lambda(E)$. $C(E)$ describes the
breakdown in the normalization of the WKB wavefunction at long range and scales
the short-range solutions to match the long-range ones. In addition the regular
and irregular WKB solutions propagated out of the short-range region lose their
phase relationship, and this loss is corrected by a phase shift given by
$\tan\lambda(E)$ \cite{Julienne:1989}. At sufficiently high energies, the WKB
wavefunctions are valid everywhere and $C(E) \rightarrow 1$ and
$\tan\lambda(E)\rightarrow 0$.

The threshold behavior of the resonance width and shift may be written in terms
of the MQDT functions, $C_{\rm bg}(E)$ and $\tan\lambda_{\rm bg}(E)$
\cite{Mies:1984a,Mies:1984},
\begin{subequations}
\begin{align}
\frac12\Gamma_n(E)=\frac12\bar{\Gamma}_n C_{\rm bg}(E)^{-2};  \\
\delta E_n(E) = -\frac12\bar{\Gamma}_n\tan\lambda_{\rm bg}(E).
\end{align}\label{mqdtsubs}
\end{subequations}
The full expression for the phase shift near a resonance is then
\begin{equation}
\eta(E,B)= \eta_{\rm bg}(E) + \eta_{\rm res}(E,B), \label{skel_etae}
\end{equation}
where
\begin{eqnarray}
\eta_{\rm res}(E,B)&&=\nonumber\\
-\tan^{-1}&&\left(\frac{\frac12\bar{\Gamma}_n C_{\rm bg}(E)^{-2}}
{E-\delta\mu(B-B_n)+\frac12\bar{\Gamma}_n\tan\lambda_{\rm bg}(E)}\right).
\label{eta_mqdt}
\end{eqnarray}

\begin{figure}[tbp]
\includegraphics[width=1\columnwidth]{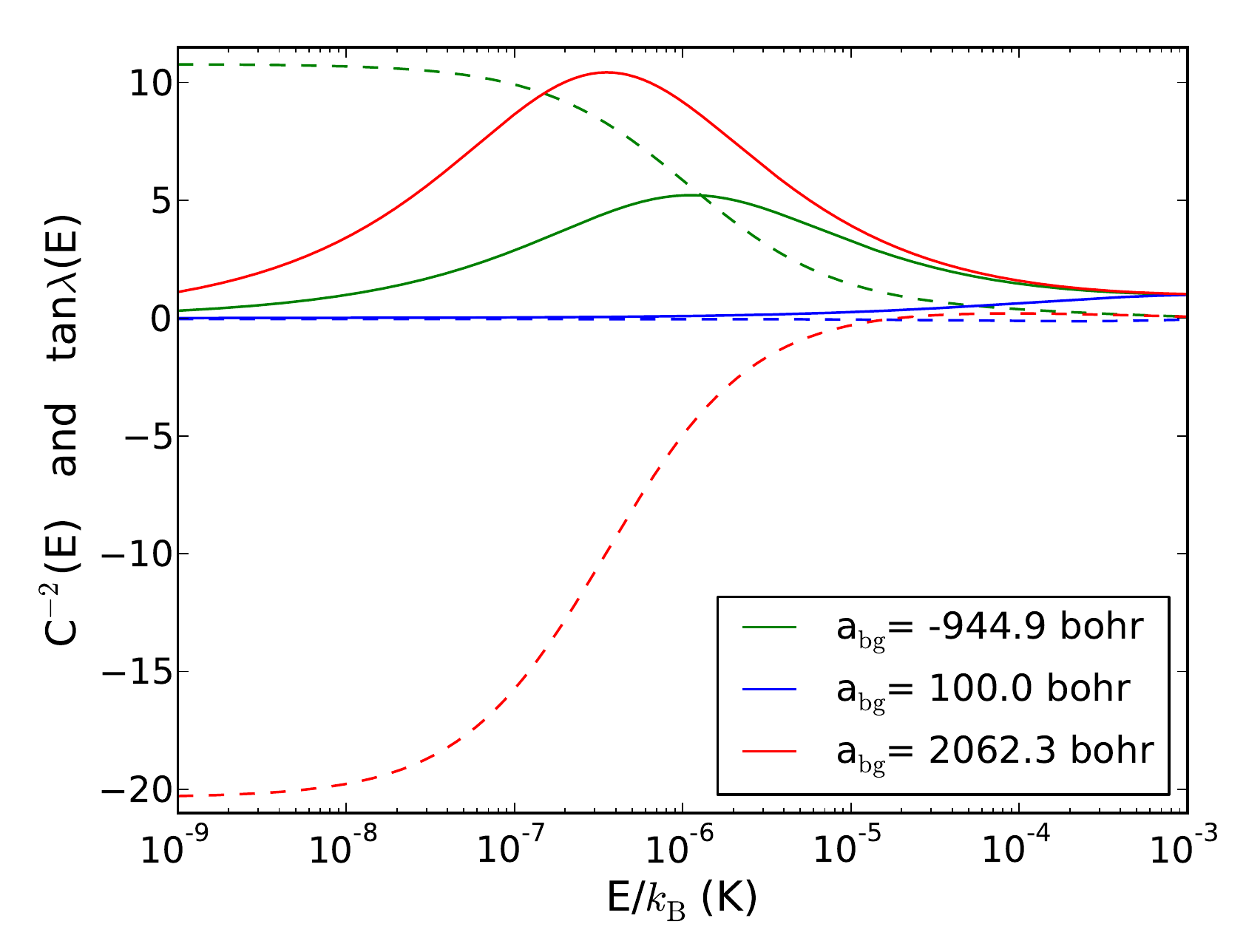}
\caption{The functions $C^{-2}(E)$ (solid) and $\tan\lambda(E)$ (dashed) for Cs,
with a variety of different background scattering lengths. The behavior of
these functions is determined by both the $C_6$ co-efficient and the background
scattering length.} \label{fig:MQDTbgparam_Cs}
\end{figure}

In the present work we follow Gao's work on analytical van der Waals theory
\cite{Gao:C6:1998, Gao:QDT:1998, Gao:bohr:1999, Gao:2000, Gao:2001, Gao:2004}
and choose reference functions that have the correct long-range $C_6$
coefficient and directly reproduce the background scattering length of the
resonance. The background phase shift $\eta_{\rm bg}(E)$ and the MQDT functions
$C^{-2}_{\rm bg}(E)$ and $\tan\lambda_{\rm bg}(E)$ are then determined
analytically by Gao's theory once the background scattering length $a_{\rm bg}$
is specified. The expression for $\eta_{\rm bg}(E)$ is given by Eq.~(2) of
Ref.~\cite{Gao:2000}, and similar expressions for the other two functions have
been derived and implemented numerically by Gao \cite{Gao:AQDTroutines}. The
energy dependence of $C^{-2}_{\rm bg}(E)$ and $\tan\lambda_{\rm bg}(E)$ for
$^{133}$Cs are shown in Fig.\ \ref{fig:MQDTbgparam_Cs} for a variety of
different $a_{\rm bg}$. The threshold behavior (accurate for $k a_{\rm bg} \ll
1$) is $C^{-2}_{\rm bg}(E) \rightarrow k\bar{a}(1+(a-r)^2)$ and
$\tan\lambda_{\rm bg} \rightarrow 1-r$ as $E \rightarrow 0$
\cite{Julienne:2006}, where $r=a_{\rm bg}/\bar{a}$. In this notation,
$\frac12\bar{\Gamma}_n$ is related to the magnetic resonance width $\Delta$ by
\begin{equation}
\frac12\bar{\Gamma}_n=\frac{r}{1+(1-r)^2}\delta\mu\Delta.
\end{equation}

To implement Eq.\ \eqref{eta_mqdt}, we first carry out coupled-channel
calculations of $a(B)$ and (if necessary) extrapolate to zero energy. We then
fit the zero-energy scattering length to Eq.\ \eqref{aB} to find the resonance
position $B_0$, magnetic field width $\Delta$ and local $a_{\rm bg}$. Along
with the van der Waals coefficient $C_6$ and the reduced mass $\mu$, this
allows us to find the MQDT parameters $C^{-2}_{\rm bg}(E)$ and
$\tan\lambda_{\rm bg}(E)$ using Gao's analytic van der Waals routines
\cite{Gao:AQDTroutines}. The shift between $B_n$, the crossing of the bare
bound state, and the coupled-channels resonance pole $B_0$ is
\begin{equation}
B_0 - B_n = \Delta r \left(\frac{1-r}{1+(1-r)^2}\right) .
\end{equation}
Lastly, we need $\delta\mu$, the difference between the magnetic moments of the
bare bound state and the separated atoms. To obtain this, we carry out
coupled-channels calculations on the near-threshold bound states of the system,
using the approach described in ref.\ \cite{Hutson:Cs2:2008}. Such calculations
give the energies of real bound states rather than bare states, but it is
usually straightforward to find a region of magnetic field where the energies
are only weakly perturbed by avoided crossings, and to obtain magnetic moments
by finite differences in this region. If necessary, pairs of crossing states
could be deperturbed to find the properties of the underlying bare states, but
this was not necessary in the present work. Typically 2-3 significant figures
were found to be sufficient in our calculations.

\section{Effectiveness of the MQDT formula}

\begin{figure}[htbp]
\centering
\includegraphics[width=1\columnwidth]{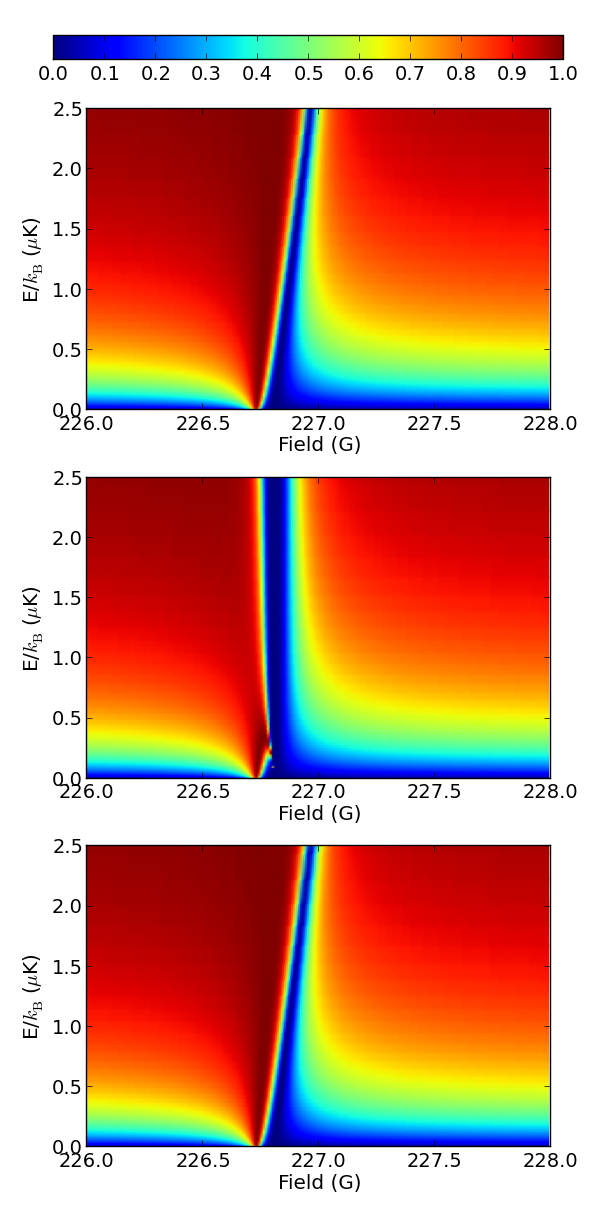}
\caption{Contour plot of $\sin^2\eta(E,B)$ for $E>0$ around the resonance at
$B_0=226.73$~G in the aa channel of $^{133}$Cs, where $E=0$ is the energy of
the two separated atoms. $\eta(E,B)$ is calculated using coupled-channels
calculations (top), the effective-range expansion, Eq.\ \eqref{r0}, (middle),
and the MQDT approach, Eq.\ \eqref{skel_etae}, (bottom).}
\label{fig:mqdtcontours}
\end{figure}

\begin{figure}[htbp]
\centering
\includegraphics[width=1\columnwidth]{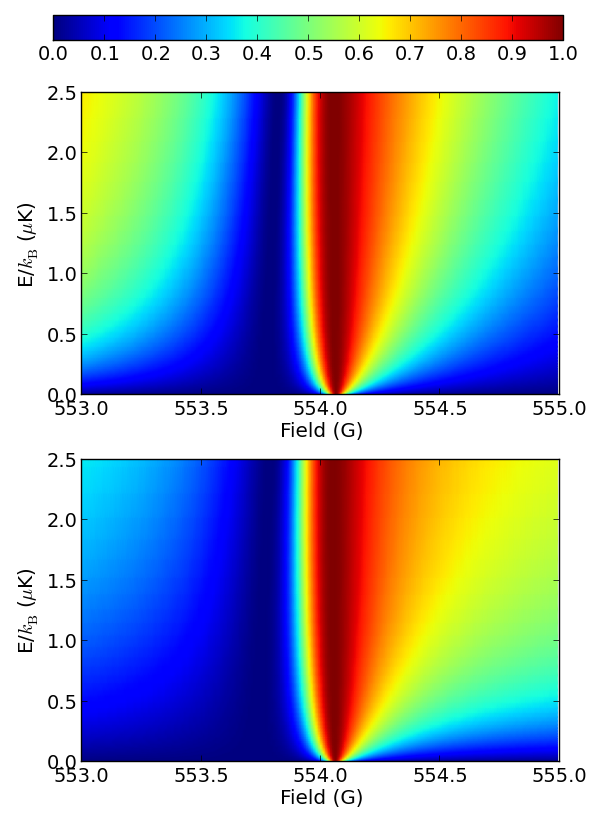}
\caption{Contour plot of $\sin^2\eta(E,B)$ for $E>0$ around the resonance at
$B_0=554$~G in the aa channel of $^{133}$Cs, where $E=0$ is the energy of
the two separated atoms. $\eta(E,B)$ is calculated using coupled-channels
calculations (top) and the MQDT approach, Eq.\ \eqref{skel_etae}, (bottom).
Around the pole of the resonance where the $a_{\rm bg}$ fitted
by the pole formula is roughly accurate then the MQDT approach works well,
towards the edges of the resonance where the $a_{\rm bg}$ is shifted the
formula breaks down.} \label{fig:Cs554G}
\end{figure}

The MQDT formula for the energy-dependent phase shift, Eq. \eqref{eta_mqdt},
was applied to the same set of narrow resonances discussed in Section
\ref{sec:eff_range_analysis}. The parameters obtained for the MQDT approach are
given in Table \ref{tab:mqdt_param}. Figure \ref{fig:phase_shift} compares the
MQDT results with those obtained directly from coupled-channels calculations at
a variety of fields around each resonance. There is excellent agreement in all
cases, and MQDT succeeds in reproducing the complicated variation of the phase
shift with both energy and field (which the effective-range expansion was
unable to do). The MQDT approach also gives results indistinguishable from
coupled-channels calculations for the broad resonance in $^6$Li, shown in Fig.\
\ref{fig:broad_reff}, although in this case the effective-range expansion is
also successful, provided the field-dependence of $r_{\rm eff}$ is taken from
coupled-channel calculations and not from an approximate formula.
\begin{table*}
\begin{tabular}{ccccccccccc}
\hline
\hline
System	\quad& $\mu~(m_{\rm u})$\quad	& $\bar{E}$~(mK)\quad	&$\bar{a}~(\hbox{bohr})$\quad&
$C_6~(E_{\rm h} \hbox{bohr}^6)$\quad& $B_0$~(G)\quad&$\Delta$~(G)\quad& $a_{\rm bg}~(\hbox{bohr})$\quad	&$B_n$~(G)\quad& $\delta\mu~(\mu_{\rm B})$ & $s_{\rm res}$ \\ \hline
$^6$Li		&3.0076	&32.3	& 29.88	& 1393.39&543.40	&	0.10		& 59.0			& 543.50	& 1.97 & $8.1\times10^{-4}$\\
$^{39}$K		&19.4819	&1.17	& 61.77	& 3926.9	&745.93	&  $-0.005$	& $-33.4$			& 744.93	& 3.95 &$6.2\times10^{-4}$\\
$^{133}$Cs	&66.4527	&0.14	& 96.62	& 6890.48&226.73	&	0.076	& 2062.26		&226.81	& 0.24 & 0.19	\\
\hline
\hline
\end{tabular}
\caption{The resonance parameters required for two-channel formula.}
\label{tab:mqdt_param}
\end{table*}

The MQDT approach can be used to generate a smooth and accurate representation
of the resonance with magnetic field both near threshold and at higher
energies. Figure \ref{fig:mqdtcontours} shows contour plots of $\sin^2\eta$ as
a function of both magnetic field and energy over the width of the Cs resonance
at 226.73~G, as obtained from coupled-channel calculations, from the effective
range expansion and from the MQDT approach. The states that arise in the
two-channel model for this resonance are shown in Fig.\
\ref{fig:twochan_model}. The shift from $B_0$ and $B_n$ between the dressed and
bare state pictures can be clearly seen. It may be seen that the MQDT approach
reproduces the coupled-channel results very accurately over the whole range of
energy and field, while the effective-range expansion does not. In particular,
the peak of the resonance, where $\sin^2\eta=1$ and $a(B)=\infty$, follows a
quite incorrect path as a function of energy in the effective-range expansion.

All the calculations described above were carried out with MQDT functions that
represent the `bare' open channel derived from the local $a_{\rm bg}$ of the
resonance, even if it is not the overall $a_{\rm bg}$ of the system. This
approach works well for the examples discussed, but it is limited to resonances
where $a_{\rm bg}$ remains reasonably constant over the width of the resonance.
This is true for most resonances with $s_{\rm res} \ll 1$, unless they sit very
close to the pole of a much wider resonance; under such circumstances, however,
there can be a substantial variation in $a_{\rm bg}$ over the width of the
resonance. As an example of this we consider the resonance at $B_0=554$~G in
the aa channel of $^{133}$Cs, which is close to the pole of a broad resonance
at 548~G.  In Fig.\ \ref{fig:Cs554G} the energy-dependent phase shift from
coupled-channels calculations is compared to the results of the MQDT approach
with fixed $a_{\rm bg}$. Whilst the MQDT approximation is good at fields close
to the pole of the resonance, it quickly starts to fail at fields further away.
This is because the two resonances need to be treated together as a pair of
interfering, overlapping resonances~\cite{Jachymski:2013} instead of treating
them as independent. In such a case, the assumption of a constant background
scattering length is valid only close to the resonance pole.

\section{Conclusion}

An accurate description of the energy dependence of the scattering phase shift
and hence the scattering length is crucial to many experiments on few-body
phenomena at finite temperatures. We have explored the behavior of the commonly
used effective-range expansion, and shown that is reasonably good at describing
the energy dependence around broad resonances and away from zero-crossings in
the scattering length. However, around narrow resonances the effective-range
expansion can fail badly, even when the full field-dependence of the effective
range is taken from coupled-channel calculations.

Gao~\cite{Gao:QDT:1998} and Flambaum {\it et al.}~\cite{Flambaum1999} have
developed an approximate formula relating the effective range $r_{\rm eff}$ to
the scattering length. We have shown that this formula is reasonably accurate
near the pole of a broad resonance, but even for broad resonances it breaks
down badly near zero-crossings, and may give an effective range of the wrong
sign. However, it is possible to write a modified form of the formula (with a
different parabolic denominator) that gives a good representation of the
effective range across the whole width of the resonance. For narrow resonances,
an analogous parabolic form may still be used, but its parameters are
completely different from those of refs.\ \cite{Gao:QDT:1998} and
\cite{Flambaum1999}.

To remedy the deficiencies of the effective-range expansion around narrow
resonances, we advocate the use of an MQDT approach that fully describes the
effect of a resonance as a function of both field and energy. This method
entails representing the resonance in a two-channel model in which a bare bound
state interacts with a bare continuum channel. The parameters of the model are
obtained from coupled-channel calculations on the bound states and scattering
length of the system. This MQDT approach successfully characterizes the
behavior of the resonance for both broad and narrow resonances. It can be
used to include the role of collision at finite energy, correct for zero-point
energy in lattices, and to evaluate thermodynamic properties of cold atoms and
molecules.

The MQDT approach described here is accurate only for individual isolated
resonances that have a reasonably constant background scattering term across
their entire width. It is not uncommon to find cases of overlapping resonances
where treating individual resonances as isolated can break down to a lesser or
greater extent.  A full treatment of overlapping resonances would require a
multi-channel treatment such as the generalized MQDT model presented by
Jachymski and Julienne \cite{Jachymski:2013}. The energy-dependent scattering
length of this model should be capable of describing the complicated variation
of the scattering phase shift with energy $E$ and magnetic field $B$ even when
there are several resonances that overlap within their widths.

Our analytic expressions for the the near-threshold energy-dependent scattering
length could benefit a number of active cold atom research areas mentioned in
the Introduction, since energies in the $\mu$K range are common with cold atom
phenomena. This could be especially important for studies of optical lattice
structures, where the finite zero-point or band energy can lead to significant
corrections to the energy of confinement-induced resonances \cite{Naidon2007}
and accounting for it requires the scattering length at finite energy
\cite{Blume2002, Bolda2002}.  Accurate finite-energy corrections to the phase
shift could also be significant for the equation of state of cold fermions
\cite{Liu:2013} and for understanding few-body phenomena \cite{Braaten2006,
WangRev2013}.

\begin{acknowledgments}
The authors are grateful to Yujun Wang for discussions. The authors acknowledge
the support of EOARD Grant FA8655-10-1-3033, AFOSR-MURI FA9550-09-1-0617, and
Engineering and Physical Sciences Research Council Grant no.\ EP/I012044/1. CLB
is supported by a Doctoral Fellowship from Durham University.
\end{acknowledgments}

\bibliography{Effective_range_extrabib,../Allrefs_psj,../all}

\end{document}